\documentclass[prb,twocolumn,showpacs,preprintnumbers,amsmath,amssymb,superscriptaddress]{revtex4-1}

\usepackage{graphicx}
\usepackage{dcolumn}
\usepackage{bm}
\usepackage{amsthm}
\usepackage{amsmath}
\usepackage{amssymb}
\usepackage{epstopdf}

\begin{document}

\title{Dipolar spin-misalignment correlations in inhomogeneous magnets: \\ comparison between neutron scattering and micromagnetic approaches}

\author{Sergey~Erokhin}\email{s.erokhin@general-numerics-rl.de}
\address{General Numerics Research Lab, An der Leite 3B, D-07749 Jena, Germany}
\author{Dmitry~Berkov}
\address{General Numerics Research Lab, An der Leite 3B, D-07749 Jena, Germany}
\author{Andreas~Michels}
\address{Physics and Materials Science Research Unit, University of Luxembourg, 162A~Avenue de la Fa\"iencerie, L-1511 Luxembourg, Grand Duchy of Luxembourg}

\begin{abstract}
In inhomogeneous bulk ferromagnets, the dominating sources of spin disorder are related to spatial variations of (i) the magnitude of the local saturation magnetization and of (ii) the magnitude and/or direction of the magnetic anisotropy field. For the particular example of a porous ferromagnet, where the magnetization inhomogeneity is at maximum, we demonstrate, by means of experimental neutron scattering data and micromagnetic simulations, the anisotropic character of magnetization fluctuations induced by the dipolar interaction.  
\end{abstract}

\maketitle\

\section{Introduction}

In polycrystalline bulk ferromagnets, \cite{herzer13} the sources of spin disorder are related to lattice imperfections, \textit{e.g.}, point defects, dislocations, grain- and phase boundaries, or pores. These microstructural defects are accompanied by spatial variations of the materials parameters, for instance, the magnitude of the local saturation magnetization, exchange constant, or variations in the magnitude and/or direction of the magnetic anisotropy field. As a result, at a given value of the applied magnetic field, these features give rise to a deviation of the magnetization from the perfectly aligned state, in other words, they result in spin misalignment. On the other hand, inhomogeneous spin states in the bulk of a material (with $\nabla \cdot \mathbf{M} \neq 0$) go along with a magnetodipolar interaction field, which has an important impact on magnetic properties.

The dipole-dipole interaction---one of the most fundamental interactions in condensed-matter physics---is still the subject of current research. For instance, it gives rise to anomalous features in the ground-state correlations and in the spin-wave excitation spectrum of 2D spin systems consisting of cold polar molecules, \cite{PhysRevLett.109.025303} and it is vital for the understanding of spin-ice physics, where frustration, dipolar ferromagnetic coupling, exotic field-induced phase transitions, and unusual glassiness are of relevance (see, \textit{e.g.}, Refs.~\onlinecite{castelnovo2012,moessnerrmp} for recent reviews). Recent experiments on PdFe islands on are square lattice using photoemission electron microscopy \cite{PhysRevLett.110.177209} even indicate that pole interactions of higher order than the dipolar one are required in order to understand the ground-state ordering features of such a system. 

While classical ``standard'' magnetometry provides only integral information about the magnetic state of the sample, scattering techniques, in particular, magnetic neutron scattering yield spatially and time-resolved information about magnetic media. A further important difference between magnetometry and magnetic neutron scattering relates to the impact of the magnetodipolar interaction: the quantity of interest in an elastic magnetic neutron scattering experiment, the differential scattering cross section $d \Sigma_M / d \Omega$, depends in a \textit{twofold} manner on the magnetodipolar interaction. First, and different from magnetization measurements, the interaction of the magnetic moment of the neutron with the sample's magnetization results in dipolar selection rules which are embodied, \textit{e.g.}, by the appearance of trigonometric functions in $d \Sigma_M / d \Omega$ (via the Halpern-Johnson vector). \cite{squires} Second, the dipole-dipole interaction between the magnetic moments in the sample has a direct impact on its magnetization structure and therefore on the Fourier components of the magnetization. The former determine the magnetization (as measured by magnetometry) and the latter the properties of $d \Sigma_M / d \Omega$. 

Traditionally, the influence of lattice defects on the magnetization of bulk magnetic materials is studied by analyzing magnetization curves in the approach-to-saturation regime. Early investigations by Brown, \cite{brown40} using the continuum theory of micromagnetics, provide analytic expressions describing different types of defects. From the neutron-scattering point of view, it is well known that the spin perturbations that are related to imperfections give rise to a strong magnetic scattering signal---the so-called spin-misalignment scattering---along the forward direction (at small scattering angles). \cite{michels2014review} The size of perturbed regions is characterized by a field-dependent correlation length, which varies between about a few nanometers up to $\sim 100 \, \mathrm{nm}$. It is the purpose of this paper to study the role of the magnetodipolar interaction on (real-space) magnetic correlations. In particular, we aim to disentangle the twofold impact of the anisotropic dipole-dipole interaction on magnetic correlation functions obtained from neutron data. 

In our analysis we employ our previously developed micromagnetic simulation methodology, which has provided fundamental insights into the magnetic small-angle neutron scattering (SANS) of various materials. \cite{erokhin2012prb,*michels2012prb1,*michels2014jmmm} The particular strength of this micromagnetic approach is that it takes into account \textit{site-dependent} magnetic interactions (for exchange, magnetic anisotropy, and saturation magnetization). This feature implies that the magnetic microstructure of a wide range of \textit{polycrystalline} magnetic materials such as single-phase nanocrystalline magnets, magnetic nanocomposites, or magnetic particles in a nonmagnetic matrix can be studied. As a prime candidate for a system with strong internal dipolar fields and nontrivial magnetization correlations, we have chosen for the present study porous ferromagnets (iron and cobalt), since here local variations in the saturation magnetization are at maximum.

\section{Micromagnetic simulation methodology}

Our micromagnetic algorithm was originally developed for the computation of the magnetization distribution of magnetic nanocomposites and of the related magnetic-field-dependent SANS cross sections. \cite{erokhin2012prb,*michels2012prb1,*michels2014jmmm} The four standard contributions to the total magnetic energy (external field, magnetic anisotropy, exchange and dipolar interaction) are taken into account. In the simulations presented here, the sample volume $V = 0.2 \times 0.75 \times 0.75 \, \mathrm{\mu m}^3$ was divided into $N \sim 5 \times 10^5$ mesh elements, comprising both pores and nanocrystallites. For the later comparison to experimental neutron data on nanocrystalline inert-gas-condensed porous iron (Ref.~\onlinecite{elmas09}) the volume fraction of pores was chosen as $P = 32 \, \%$, with randomly placed pore centers. Due to the flexibility of the mesh-generation algorithm, the shape of the pores can be controlled and was taken to be polyhedron-like. The pore-size distribution was assumed to be lognormal \cite{krill98} with a median of $15 \, \mathrm{nm}$ and a variance of $1.16$, which yields a maximum of the distribution at $12 \, \mathrm{nm}$. The \textit{local} saturation magnetization of each (iron) nanocrystallite was taken $\mu_0 M_s = 2.2 \, \mathrm{T}$, which in conjunction with the above mentioned porosity value yields $\mu_0 \overline{M_s} \cong 1.5 \, \mathrm{T}$ for the entire sample. For the exchange-stiffness constant and the first cubic anisotropy constant of iron, we have, respectively, assumed values of $A = 25 \, \mathrm{pJ/m}$ and $K_1 = 47 \, \mathrm{kJ/m^3}$ (Ref.~\onlinecite{cullitygraham05}). The direction of anisotropy axes varies randomly from crystallite to crystallite. The energy-minimization procedure provides (at some particular value of the applied magnetic field) the magnetization vector field $\mathbf{M}(\mathbf{r}) = [M_x(\mathbf{r}), M_y(\mathbf{r}), M_z(\mathbf{r})]$ of the sample on an \textit{irregular} lattice. This distribution is then mapped onto a \textit{regular} lattice, which permits us to calculate the magnetization Fourier coefficients and the ensuing neutron scattering cross section using Fast Fourier transformation. Further details can be found in Refs.~\onlinecite{erokhin2012prb,*michels2012prb1,*michels2014jmmm}.

\section{Results and discussion}

\begin{figure}[tb]
\centering
\resizebox{1.0\columnwidth}{!}{\includegraphics{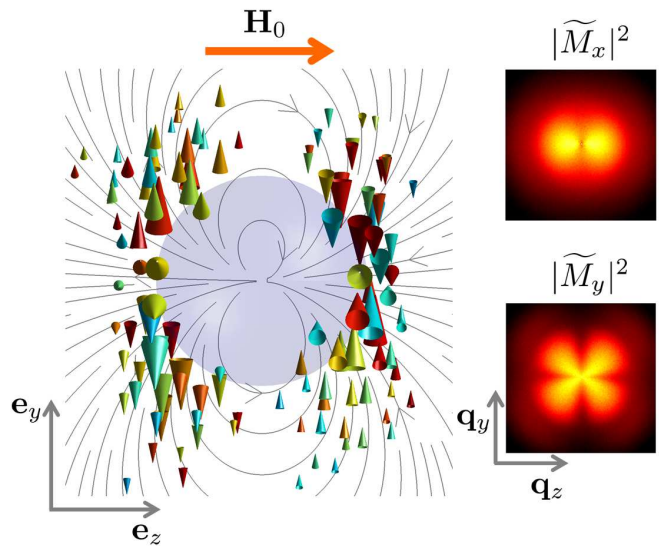}}
\caption{(Color online) (left) Computed spin misalignment (at $\mu_0 H_0 = 0.6 \, \mathrm{T}$) around a pore ($D = 12 \, \mathrm{nm}$) in a ferromagnetic iron matrix (2D cut out of a 3D simulation). Shown is the magnetization component $\mathbf{M}_\perp(\mathbf{r})$ perpendicular to $\mathbf{H}_0 \parallel \mathbf{e}_z$; thickness of arrows is proportional to the magnitude of $\mathbf{M}_\perp$. Solid grey lines: magnetodipolar field distribution. (right) Corresponding magnetization Fourier components $|\widetilde{M}_x(\mathbf{q})|^2$ and $|\widetilde{M}_y(\mathbf{q})|^2$ projected into the plane $q_x = 0$. Bright colors correspond to ``high'' values and dark colors to ``low'' values of the Fourier components. Pixels in the corners of the images have $q \cong 0.4 \, \mathrm{nm}^{-1}$. Logarithmic color scale is used.}
\label{fig1}
\end{figure}

Figure~\ref{fig1} displays the numerically computed spin misalignment $\mathbf{M}_\perp(\mathbf{r}) = (M_x, M_y, 0)$ in the vicinity of a nanopore together with the transversal Fourier coefficients $|\widetilde{M}_x(\mathbf{q})|^2$ and $|\widetilde{M}_y(\mathbf{q})|^2$; these Fourier components will be used later on when discussing magnetic neutron scattering. The discontinuity (jump) of the magnetization at the pore-matrix interphase ($\mu_0 \Delta M = 2.2 \, \mathrm{T}$) gives rise to a strong magnetostatic field and to the characteristic dipole-field-type spin texture. The symmetry of the $\mathbf{M}_\perp(\mathbf{r})$ distribution corresponds to the field of a magnetic dipole located at the center of the pore and aligned opposite to the external field direction, as depicted by the solid lines in Fig.~\ref{fig1}. Both transversal Fourier components are highly anisotropic, whereas the longitudinal one $|\widetilde{M}_z|^2$ is isotropic (data not shown). We would like to emphasize that the angular anisotropy of both $|\widetilde{M}_x|^2$ and $|\widetilde{M}_y|^2$ is a consequence of the internal magnetodipolar field, which is the only long-range (nonlocal) anisotropic interaction in our micromagnetic modeling; neglecting this energy term results in $|\widetilde{M}_x|^2$ and $|\widetilde{M}_y|^2$ being isotropic. \cite{erokhin2012prb,*michels2012prb1,*michels2014jmmm}

However, in contrast to our previous results, \cite{erokhin2012prb,*michels2012prb1,*michels2014jmmm} where $|\widetilde{M}_x|^2$ was found to be \textit{isotropic}, the present simulations reveal an \textit{anisotropic} $|\widetilde{M}_x|^2$ Fourier coefficient (which is enhanced along the field direction). This finding is due to a corrected averaging procedure, which takes into account magnetic fluctuations also along the $\mathbf{e}_x$ direction, which coincides with the direction of the incident neutron beam in a SANS experiment (see Fig.~\ref{fig1}). Specifically, the computed (mapped on a regular grid) spin structure of the sample, $\mathbf{M}(x, y, z)$, is divided into thin slices $i = 1, ..., N_x$ (with a typical thickness of $2.5 \, \mathrm{nm}$). This results in a set of magnetization distributions $\mathbf{M}^{(i)}(y, z)$, and Fourier transformation then yields the $\widetilde{\mathbf{M}}^{(i)}(q_y, q_z)$. The squared Fourier coefficients $|\widetilde{M}^i_x|^2$ etc.\ are summed up and divided by the number of slices in order to obtain the averaged quantities.

As mentioned above, the possibility to compute the individual Fourier components allows one to compare the simulation results with experimental data for the spin-misalignment SANS cross section, which reads (for the scattering geometry where $\mathbf{H}_0 \parallel \mathbf{e}_z$ is applied perpendicular to the incoming neutron beam): \cite{michels2014review}
\begin{eqnarray}
\label{sigmasansperp}
\frac{d \Sigma_M}{d \Omega}(\mathbf{q}) = \frac{8 \pi^3}{V} \, b_H^2 \left( |\widetilde{M}_x|^2 + |\widetilde{M}_y|^2 \cos^2\theta \right. \nonumber \\ \left. - (\widetilde{M}_y \widetilde{M}_z^{\ast} + \widetilde{M}_y^{\ast} \widetilde{M}_z) \sin\theta \cos\theta \right) ,
\end{eqnarray}
where $V$ is the scattering volume, $b_H = 2.9 \times 10^{8} \mathrm{A^{-1} m^{-1}}$, and $\theta$ denotes the angle between the momentum-transfer vector $\mathbf{q} \cong (0, q_y, q_z)$ and $\mathbf{e}_z$ ($a^{\ast}$ is the complex conjugate of $a$). Note that $d \Sigma_M / d \Omega$ represents the part of the total unpolarized SANS cross section $d \Sigma / d \Omega$ which is exclusively due to transversal spin misalignment, with corresponding Fourier amplitudes $\widetilde{M}_x$ and $\widetilde{M}_y$; in other words, $d \Sigma_M / d \Omega$ is obtained by subtracting the nuclear and magnetic SANS at complete saturation, $d \Sigma / d \Omega = \frac{8 \pi^3}{V} (|\widetilde{N}|^2 + b^2_H |\widetilde{M}_z|^2 \sin^2\theta)$, from the measured $d \Sigma / d \Omega$ at lower fields. 

The trigonometric functions and the $\widetilde{M}_y \widetilde{M}_z$-containing cross term in Eq.~(\ref{sigmasansperp}) are due to the neutron-magnetic interaction, while the Fourier components $\widetilde{M}_{x,y,z}$ may additionally depend on the internal magnetodipolar interaction, \textit{e.g.}, the anisotropy of the dipolar interaction is embodied in the dependence of the $\widetilde{M}_{x,y}$ on the angle $\theta$ (see Fig.~\ref{fig1}). Due to the complexity of this expression, the Fourier transform of $d \Sigma_M / d \Omega$, which one may call the correlation function of the spin-misalignment SANS cross section, does not represent, of course, the autocorrelation function of the magnetization. This is in contrast to the well-known result from nuclear scattering, where the nuclear scattering cross section
\begin{equation}
\label{sigmanuc}
\frac{d \Sigma_N}{d \Omega}(\mathbf{q}) \sim \int C_N(\mathbf{r}) \, \exp(- i \mathbf{q} \mathbf{r}) \, d^3r
\end{equation}
is equal to the Fourier transform of the autocorrelation function of the nuclear density: \cite{fournet,*porod,*feigin}
\begin{equation}
\label{cnucdef}
C_N(\mathbf{r}) \sim \int \Delta N(\mathbf{x}) \, \Delta N(\mathbf{x + r}) \, d^3x ,
\end{equation}
where $\Delta N(\mathbf{r}) = N(\mathbf{r}) - \langle N \rangle$ denotes the so-called excess scattering-length density, and $\langle N \rangle$ is the (constant) average scattering-length density, which only gives a contribution to $d \Sigma_N / d \Omega$ at $\mathbf{q} = 0$.

In order to compare the results for the correlation function of the spin-misalignment SANS cross section,
\begin{equation}
\label{cmdef}
C_M(y, z) \sim \int \frac{d \Sigma_M}{d \Omega}(q_y, q_z) \, \exp(i \mathbf{q} \mathbf{r}) \, d^2q ,
\end{equation}
with the autocorrelation function of the magnetization spin misalignment, $C_{SM}(\mathbf{r})$, which is not decorated by the neutron-magnetic interaction, we define the latter as follows: \cite{michels03prl,*weissm04a,*dobrichprb2012}
\begin{eqnarray}
\label{csmdef}
C_{SM}(\mathbf{r}) &\sim& \int \mathbf{M}_\perp(\mathbf{x}) \, \mathbf{M}_\perp(\mathbf{x+r}) \, d^3x.
\end{eqnarray}
Using the convolution theorem, this expression can be rewritten as
\begin{eqnarray}
\label{csmconvol}
C_{SM}(\mathbf{r})  &\sim& \int \left( |\widetilde{M}_x(\mathbf{q})|^2 + |\widetilde{M}_y(\mathbf{q})|^2 \right) \, \exp(i \mathbf{q} \mathbf{r}) \, d^3q .
\end{eqnarray}
Note the additional difference between both correlation functions: $C_M$ involves only a 2D integration over $\mathbf{q}$ space, because $d \Sigma_M / d \Omega$ is experimentally accessible only in the plane perpendicular to the incoming neutron beam [small-angle approximation: $\mathbf{q} \cong (0, q_y, q_z)$], whereas $C_{SM}$ is naturally obtained by a 3D Fourier transformation [due to $\widetilde{M}_{x, y} = \widetilde{M}_{x, y}(q_x, q_y, q_z)]$. Therefore, in order to compare both correlation functions, we have to employ for the computation of $C_{SM}$ the \textit{same} averaged 2D Fourier components $|\widetilde{M}_x|^2$ and $|\widetilde{M}_y|^2$ that enter $d \Sigma_M / d \Omega$, \textit{i.e.}, the integration in Eq.~(\ref{csmdef}) is also only carried out over the $y-z$ plane.

\begin{figure}[tb]
\centering
\resizebox{1.0\columnwidth}{!}{\includegraphics{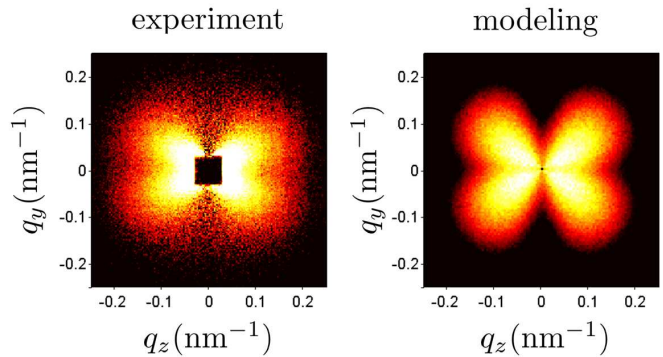}}
\caption{(Color online) Semiquantitative comparison between experimental data \cite{elmas09} and micromagnetic simulations for the spin-misalignment SANS cross section of nanoporous iron (porosity: $P = 32 \, \%$; $\mu_0 H_0 = 0.6 \, \mathrm{T}$). The respective scattering signal at a saturating field of $1.83 \, \mathrm{T}$ has been subtracted in both data sets. Logarithmic color scale is used.}
\label{fig2}
\end{figure}

Before discussing the correlations functions, we show in Fig.~\ref{fig2} a comparison between the experimental (spin-misalignment) SANS cross section of nanoporous iron \cite{elmas09} and the results of the micromagnetic simulations. The simulation data exhibit an overall good agreement with the experimental data and the anisotropic character of $d \Sigma_M / d \Omega$ is clearly reproduced; this supports the validity of our micromagnetic simulation methodology. 

\begin{figure}[tb]
\centering
\resizebox{1.0\columnwidth}{!}{\includegraphics{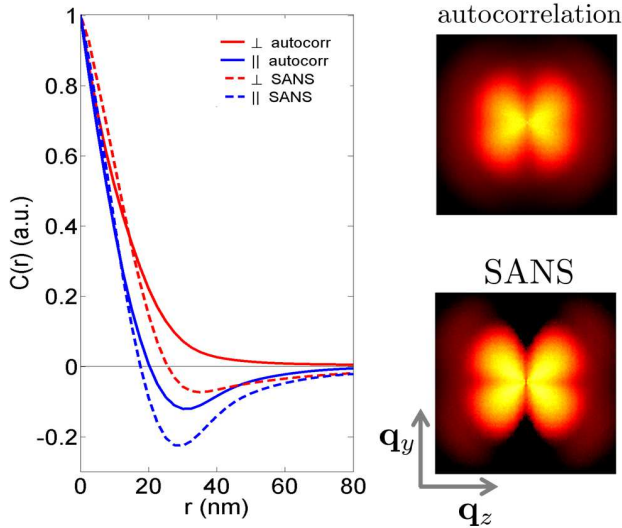}}
\caption{(Color online) Comparison between the normalized autocorrelation function of the spin misalignment $C_{SM}$ (solid lines) and the normalized correlation function of the spin-misalignment SANS cross section $C_M$ (dashed lines) along different directions in the $y-z$ detector plane ($\mu_0 H_0 = 0.6 \, \mathrm{T}$). The right images show the corresponding combination of Fourier components, projected into the plane of the 2D detector: (autocorrelation) $|\widetilde{M}_x|^2 + |\widetilde{M}_y|^2$; (SANS) $|\widetilde{M}_x|^2 + |\widetilde{M}_y|^2 \cos^2\theta - (\widetilde{M}_y \widetilde{M}_z^{\ast} + \widetilde{M}_y^{\ast} \widetilde{M}_z) \sin\theta \cos\theta$. Pixels in the corners of the images have $q \cong 0.4 \, \mathrm{nm}^{-1}$. Logarithmic color scale is used.}
\label{fig3}
\end{figure}

The correlation functions at $0.6 \, \mathrm{T}$, obtained using Eqs.~(\ref{cmdef}) and (\ref{csmconvol}), are depicted in Fig.~\ref{fig3} along the horizontal ($z$) and vertical ($y$) direction. One recognizes the existence of anisotropic correlations already for the autocorrelation function of the spin misalignment (not influenced by the interaction between neutrons and magnetic moments), which may be expected due to the long-range and anisotropic nature of the magnetodipolar interaction (see also Fig.~\ref{fig1}); the difference between both directions is significant (in particular for $r \cong 30 - 40 \, \mathrm{nm}$) with $C_{SM}$ along the vertical direction being exclusively positive definite, while $C_{SM}$ along the horizontal direction intersects the $r$-axis at $r \cong 20 \, \mathrm{nm}$ and possesses a global minimum at $r \cong 30 \, \mathrm{nm}$. The existence of anticorrelations in $C_{SM}$ around the particular $r$-value is a manifestation of the typical magnetization distribution $\mathbf{M}_\perp(\mathbf{r})$ around a pore (see Fig.~\ref{fig1}), which is due to the configuration of the magnetodipolar field in the vicinity of such an inclusion. Namely, the perpendicular magnetization component changes its sign along the direction of applied field at a distance comparable with the pore diameter. Of course, the zeros and global minima of the correlation functions are dependent on the applied magnetic field. We point out (see also explanations above) that the difference between the correlation functions corresponding to both approaches---neutron scattering and micromagnetics---is also significant, in particular, for $r \cong 20 - 50 \, \mathrm{nm}$.

\begin{figure}[]
\centering
\resizebox{1.0\columnwidth}{!}{\includegraphics{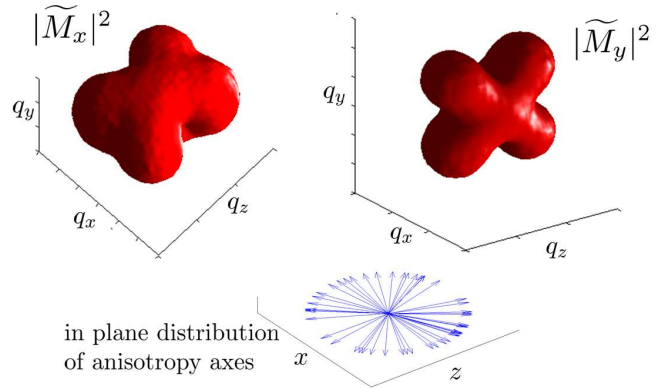}}
\caption{(Color online) Isosurfaces of the 3D magnetization Fourier components $|\widetilde{M}_x(\mathbf{q})|^2$ and $|\widetilde{M}_y(\mathbf{q})|^2$ of porous cobalt at an external magnetic field of $\mu_0 H_0 = 0.6 \, \mathrm{T}$ applied in $z$-direction. The image below schematically shows the assumed distribution of uniaxial anisotropy axes in the $x-z$ plane.}
\label{fig4}
\end{figure}

The situation becomes even more interesting if the magnetic system possesses a second symmetry breaking (in addition to the symmetry breaking caused by the external magnetic field). It can lead to the formation of \textit{nontrivial 3D correlations}, \textit{i.e.}, the correlation lengths of the magnetization distribution can be significantly different along the three coordinate axes. One can observe this effect, \textit{e.g.}, by introducing an antisymmetric exchange contribution to the total energy, also called Dzyaloshinskii-Moriya interaction, favoring the formation of spin canting. \cite{wiesendanger2015}

Another example for a magnetic system with nontrivial 3D correlations is a polycrystalline magnet with a nonisotropic (\textit{e.g.}, in-plane) distribution of anisotropy axes of the individual grains (see Fig.~\ref{fig4}). Here, we have simulated this situation for the case of nanoporous cobalt. The choice of this material was motivated by the strong \textit{uniaxial} anisotropy [$K_u = 400 \, \mathrm{kJ/m^3}$ (Ref.~\onlinecite{cullitygraham05})]. The random 2D distribution of anisotropy axes is in the $x-z$-plane. All structural parameters including the porosity value are the same as we used for the simulations on porous iron. The \textit{local} saturation magnetization and the exchange-stiffness constant of cobalt were, respectively, taken as $\mu_0 M_s = 1.76 \, \mathrm{T}$ and $A = 28 \, \mathrm{pJ/m}$ (Ref.~\onlinecite{michels00b}).

The simulation results are displayed in Fig.~\ref{fig4} as isosurfaces of the 3D magnetization Fourier components $|\widetilde{M}_x(q_x, q_y, q_z)|^2$ and $|\widetilde{M}_y(q_x, q_y, q_z)|^2$.  One can clearly see that both Fourier components are strongly influenced by the magnetodipolar interaction, which results in a 3D clover-leaf-shaped pattern that was already presented in Fig.~\ref{fig1} as a projection into the plane $q_x = 0$. But the most important observation is that $|\widetilde{M}_x|^2$ and $|\widetilde{M}_y|^2$ are qualitatively different (for a direct qualitative comparison one should rotate, \textit{e.g.}, the $|\widetilde{M}_x|^2$ component around $\mathbf{q}_z$ taking into account the relation between the Fourier components and the coordinate system). The magnetodipolar ``structure'' in $|\widetilde{M}_y|^2$ is more pronounced, since the corresponding coordinate vector $\mathbf{e}_y$ is perpendicular to the uniaxial anisotropy axes in the sample. This result shows that the magnetic correlations in porous cobalt with an in-plane distribution of anisotropy axes have a nontrivial 3D character. The components can not be fully matched by rotation around $\mathbf{q}_z$, in contrast to the case of a 3D random distribution of anisotropy axes (see the results on porous iron above).

\section{Summary and Conclusions}

Using micromagnetic simulations we have computed the magnetization distribution of nanoporous ferromagnets (iron and cobalt) which exhibits strong dipolar correlations due to nanoscale spatial variations of the magnitude of the magnetization. The results were used in order to compute the correlation function of the spin-misalignment SANS cross section, which is affected in a twofold manner by the magnetodipolar interaction, and the autocorrelation function of the spin misalignment, which by contrast depends only on the magnetization distribution. Our approach was validated by comparing the simulation results with experimental neutron data. As a consequence of the long-range and anisotropic character of the internal magnetodipolar field, we find strongly \textit{anisotropic} magnetic correlations. Our results demonstrate the importance of the magnetodipolar interaction for understanding magnetic neutron scattering.

\section*{Acknowledgements}

We thank the FNR (Project No.~INTER/DFG/12/07) and the DFG (Project No.~BE 2464/10-3) for financial support.


\begin{thebibliography}{22}%
\makeatletter
\providecommand \@ifxundefined [1]{%
 \@ifx{#1\undefined}
}%
\providecommand \@ifnum [1]{%
 \ifnum #1\expandafter \@firstoftwo
 \else \expandafter \@secondoftwo
 \fi
}%
\providecommand \@ifx [1]{%
 \ifx #1\expandafter \@firstoftwo
 \else \expandafter \@secondoftwo
 \fi
}%
\providecommand \natexlab [1]{#1}%
\providecommand \enquote  [1]{``#1''}%
\providecommand \bibnamefont  [1]{#1}%
\providecommand \bibfnamefont [1]{#1}%
\providecommand \citenamefont [1]{#1}%
\providecommand \href@noop [0]{\@secondoftwo}%
\providecommand \href [0]{\begingroup \@sanitize@url \@href}%
\providecommand \@href[1]{\@@startlink{#1}\@@href}%
\providecommand \@@href[1]{\endgroup#1\@@endlink}%
\providecommand \@sanitize@url [0]{\catcode `\\12\catcode `\$12\catcode
  `\&12\catcode `\#12\catcode `\^12\catcode `\_12\catcode `\%12\relax}%
\providecommand \@@startlink[1]{}%
\providecommand \@@endlink[0]{}%
\providecommand \url  [0]{\begingroup\@sanitize@url \@url }%
\providecommand \@url [1]{\endgroup\@href {#1}{\urlprefix }}%
\providecommand \urlprefix  [0]{URL }%
\providecommand \Eprint [0]{\href }%
\providecommand \doibase [0]{http://dx.doi.org/}%
\providecommand \selectlanguage [0]{\@gobble}%
\providecommand \bibinfo  [0]{\@secondoftwo}%
\providecommand \bibfield  [0]{\@secondoftwo}%
\providecommand \translation [1]{[#1]}%
\providecommand \BibitemOpen [0]{}%
\providecommand \bibitemStop [0]{}%
\providecommand \bibitemNoStop [0]{.\EOS\space}%
\providecommand \EOS [0]{\spacefactor3000\relax}%
\providecommand \BibitemShut  [1]{\csname bibitem#1\endcsname}%
\let\auto@bib@innerbib\@empty
\bibitem [{\citenamefont {Herzer}(2013)}]{herzer13}%
  \BibitemOpen
  \bibfield  {author} {\bibinfo {author} {\bibfnamefont {G.}~\bibnamefont
  {Herzer}},\ }\href@noop {} {\bibfield  {journal} {\bibinfo  {journal} {Acta
  Mater.}\ }\textbf {\bibinfo {volume} {61}},\ \bibinfo {pages} {718} (\bibinfo
  {year} {2013})}\BibitemShut {NoStop}%
\bibitem [{\citenamefont {Peter}\ \emph {et~al.}(2012)\citenamefont {Peter},
  \citenamefont {M\"uller}, \citenamefont {Wessel},\ and\ \citenamefont
  {B\"uchler}}]{PhysRevLett.109.025303}%
  \BibitemOpen
  \bibfield  {author} {\bibinfo {author} {\bibfnamefont {D.}~\bibnamefont
  {Peter}}, \bibinfo {author} {\bibfnamefont {S.}~\bibnamefont {M\"uller}},
  \bibinfo {author} {\bibfnamefont {S.}~\bibnamefont {Wessel}}, \ and\ \bibinfo
  {author} {\bibfnamefont {H.~P.}\ \bibnamefont {B\"uchler}},\ }\href@noop {}
  {\bibfield  {journal} {\bibinfo  {journal} {Phys. Rev. Lett.}\ }\textbf
  {\bibinfo {volume} {109}},\ \bibinfo {pages} {025303} (\bibinfo {year}
  {2012})}\BibitemShut {NoStop}%
\bibitem [{\citenamefont {Castelnovo}\ \emph {et~al.}(2012)\citenamefont
  {Castelnovo}, \citenamefont {Moessner},\ and\ \citenamefont
  {Sondhi}}]{castelnovo2012}%
  \BibitemOpen
  \bibfield  {author} {\bibinfo {author} {\bibfnamefont {C.}~\bibnamefont
  {Castelnovo}}, \bibinfo {author} {\bibfnamefont {R.}~\bibnamefont
  {Moessner}}, \ and\ \bibinfo {author} {\bibfnamefont {S.~L.}\ \bibnamefont
  {Sondhi}},\ }\href@noop {} {\bibfield  {journal} {\bibinfo  {journal} {Annu.
  Rev. Condens. Matter Phys.}\ }\textbf {\bibinfo {volume} {3}},\ \bibinfo
  {pages} {35} (\bibinfo {year} {2012})}\BibitemShut {NoStop}%
\bibitem [{\citenamefont {Nisoli}\ \emph {et~al.}(2013)\citenamefont {Nisoli},
  \citenamefont {Moessner},\ and\ \citenamefont {Schiffer}}]{moessnerrmp}%
  \BibitemOpen
  \bibfield  {author} {\bibinfo {author} {\bibfnamefont {C.}~\bibnamefont
  {Nisoli}}, \bibinfo {author} {\bibfnamefont {R.}~\bibnamefont {Moessner}}, \
  and\ \bibinfo {author} {\bibfnamefont {P.}~\bibnamefont {Schiffer}},\ }\href
  {\doibase 10.1103/RevModPhys.85.1473} {\bibfield  {journal} {\bibinfo
  {journal} {Rev. Mod. Phys.}\ }\textbf {\bibinfo {volume} {85}},\ \bibinfo
  {pages} {1473} (\bibinfo {year} {2013})}\BibitemShut {NoStop}%
\bibitem [{\citenamefont {Ewerlin}\ \emph {et~al.}(2013)\citenamefont
  {Ewerlin}, \citenamefont {Demirbas}, \citenamefont {Br\"ussing},
  \citenamefont {Petracic}, \citenamefont {\"Unal}, \citenamefont {Valencia},
  \citenamefont {Kronast},\ and\ \citenamefont
  {Zabel}}]{PhysRevLett.110.177209}%
  \BibitemOpen
  \bibfield  {author} {\bibinfo {author} {\bibfnamefont {M.}~\bibnamefont
  {Ewerlin}}, \bibinfo {author} {\bibfnamefont {D.}~\bibnamefont {Demirbas}},
  \bibinfo {author} {\bibfnamefont {F.}~\bibnamefont {Br\"ussing}}, \bibinfo
  {author} {\bibfnamefont {O.}~\bibnamefont {Petracic}}, \bibinfo {author}
  {\bibfnamefont {A.~A.}\ \bibnamefont {\"Unal}}, \bibinfo {author}
  {\bibfnamefont {S.}~\bibnamefont {Valencia}}, \bibinfo {author}
  {\bibfnamefont {F.}~\bibnamefont {Kronast}}, \ and\ \bibinfo {author}
  {\bibfnamefont {H.}~\bibnamefont {Zabel}},\ }\href@noop {} {\bibfield
  {journal} {\bibinfo  {journal} {Phys. Rev. Lett.}\ }\textbf {\bibinfo
  {volume} {110}},\ \bibinfo {pages} {177209} (\bibinfo {year}
  {2013})}\BibitemShut {NoStop}%
\bibitem [{\citenamefont {Squires}(1978)}]{squires}%
  \BibitemOpen
  \bibfield  {author} {\bibinfo {author} {\bibfnamefont {G.~L.}\ \bibnamefont
  {Squires}},\ }\href@noop {} {\emph {\bibinfo {title} {Introduction to the
  Theory of Thermal Neutron Scattering}}}\ (\bibinfo  {publisher} {Dover
  Publications},\ \bibinfo {address} {New York},\ \bibinfo {year}
  {1978})\BibitemShut {NoStop}%
\bibitem [{\citenamefont {Brown~Jr.}(1940)}]{brown40}%
  \BibitemOpen
  \bibfield  {author} {\bibinfo {author} {\bibfnamefont {W.~F.}\ \bibnamefont
  {Brown~Jr.}},\ }\href@noop {} {\bibfield  {journal} {\bibinfo  {journal}
  {Phys. Rev.}\ }\textbf {\bibinfo {volume} {58}},\ \bibinfo {pages} {736}
  (\bibinfo {year} {1940})}\BibitemShut {NoStop}%
\bibitem [{\citenamefont {Michels}(2014)}]{michels2014review}%
  \BibitemOpen
  \bibfield  {author} {\bibinfo {author} {\bibfnamefont {A.}~\bibnamefont
  {Michels}},\ }\href@noop {} {\bibfield  {journal} {\bibinfo  {journal} {J.
  Phys.: Condens. Matter}\ }\textbf {\bibinfo {volume} {26}},\ \bibinfo {pages}
  {383201} (\bibinfo {year} {2014})}\BibitemShut {NoStop}%
\bibitem [{\citenamefont {Erokhin}\ \emph
  {et~al.}(2012{\natexlab{a}})\citenamefont {Erokhin}, \citenamefont {Berkov},
  \citenamefont {Gorn},\ and\ \citenamefont {Michels}}]{erokhin2012prb}%
  \BibitemOpen
  \bibfield  {author} {\bibinfo {author} {\bibfnamefont {S.}~\bibnamefont
  {Erokhin}}, \bibinfo {author} {\bibfnamefont {D.}~\bibnamefont {Berkov}},
  \bibinfo {author} {\bibfnamefont {N.}~\bibnamefont {Gorn}}, \ and\ \bibinfo
  {author} {\bibfnamefont {A.}~\bibnamefont {Michels}},\ }\href@noop {}
  {\bibfield  {journal} {\bibinfo  {journal} {Phys. Rev. B}\ }\textbf {\bibinfo
  {volume} {85}},\ \bibinfo {pages} {024410} (\bibinfo {year}
  {2012}{\natexlab{a}})}\BibitemShut {NoStop}%
\bibitem [{\citenamefont {Erokhin}\ \emph
  {et~al.}(2012{\natexlab{b}})\citenamefont {Erokhin}, \citenamefont {Berkov},
  \citenamefont {Gorn},\ and\ \citenamefont {Michels}}]{michels2012prb1}%
  \BibitemOpen
  \bibfield  {author} {\bibinfo {author} {\bibfnamefont {S.}~\bibnamefont
  {Erokhin}}, \bibinfo {author} {\bibfnamefont {D.}~\bibnamefont {Berkov}},
  \bibinfo {author} {\bibfnamefont {N.}~\bibnamefont {Gorn}}, \ and\ \bibinfo
  {author} {\bibfnamefont {A.}~\bibnamefont {Michels}},\ }\href@noop {}
  {\bibfield  {journal} {\bibinfo  {journal} {Phys. Rev. B}\ }\textbf {\bibinfo
  {volume} {85}},\ \bibinfo {pages} {134418} (\bibinfo {year}
  {2012}{\natexlab{b}})}\BibitemShut {NoStop}%
\bibitem [{\citenamefont {Michels}\ \emph {et~al.}(2014)\citenamefont
  {Michels}, \citenamefont {Erokhin}, \citenamefont {Berkov},\ and\
  \citenamefont {Gorn}}]{michels2014jmmm}%
  \BibitemOpen
  \bibfield  {author} {\bibinfo {author} {\bibfnamefont {A.}~\bibnamefont
  {Michels}}, \bibinfo {author} {\bibfnamefont {S.}~\bibnamefont {Erokhin}},
  \bibinfo {author} {\bibfnamefont {D.}~\bibnamefont {Berkov}}, \ and\ \bibinfo
  {author} {\bibfnamefont {N.}~\bibnamefont {Gorn}},\ }\href@noop {} {\bibfield
   {journal} {\bibinfo  {journal} {J. Magn. Magn. Mater.}\ }\textbf {\bibinfo
  {volume} {350}},\ \bibinfo {pages} {55} (\bibinfo {year} {2014})}\BibitemShut
  {NoStop}%
\bibitem [{\citenamefont {Michels}\ \emph {et~al.}(2009)\citenamefont
  {Michels}, \citenamefont {Elmas}, \citenamefont {D\"obrich}, \citenamefont
  {Ames}, \citenamefont {Markmann}, \citenamefont {Sharp}, \citenamefont
  {Eckerlebe}, \citenamefont {Kohlbrecher},\ and\ \citenamefont
  {Birringer}}]{elmas09}%
  \BibitemOpen
  \bibfield  {author} {\bibinfo {author} {\bibfnamefont {A.}~\bibnamefont
  {Michels}}, \bibinfo {author} {\bibfnamefont {M.}~\bibnamefont {Elmas}},
  \bibinfo {author} {\bibfnamefont {F.}~\bibnamefont {D\"obrich}}, \bibinfo
  {author} {\bibfnamefont {M.}~\bibnamefont {Ames}}, \bibinfo {author}
  {\bibfnamefont {J.}~\bibnamefont {Markmann}}, \bibinfo {author}
  {\bibfnamefont {M.}~\bibnamefont {Sharp}}, \bibinfo {author} {\bibfnamefont
  {H.}~\bibnamefont {Eckerlebe}}, \bibinfo {author} {\bibfnamefont
  {J.}~\bibnamefont {Kohlbrecher}}, \ and\ \bibinfo {author} {\bibfnamefont
  {R.}~\bibnamefont {Birringer}},\ }\href@noop {} {\bibfield  {journal}
  {\bibinfo  {journal} {EPL}\ }\textbf {\bibinfo {volume} {85}},\ \bibinfo
  {pages} {47003} (\bibinfo {year} {2009})}\BibitemShut {NoStop}%
\bibitem [{\citenamefont {Krill}\ and\ \citenamefont
  {Birringer}(1998)}]{krill98}%
  \BibitemOpen
  \bibfield  {author} {\bibinfo {author} {\bibfnamefont {C.~E.}\ \bibnamefont
  {Krill}}\ and\ \bibinfo {author} {\bibfnamefont {R.}~\bibnamefont
  {Birringer}},\ }\href@noop {} {\bibfield  {journal} {\bibinfo  {journal}
  {Philos. Mag. A}\ }\textbf {\bibinfo {volume} {77}},\ \bibinfo {pages} {621}
  (\bibinfo {year} {1998})}\BibitemShut {NoStop}%
\bibitem [{\citenamefont {Cullity}\ and\ \citenamefont
  {Graham}(2005)}]{cullitygraham05}%
  \BibitemOpen
  \bibfield  {author} {\bibinfo {author} {\bibfnamefont {B.~D.}\ \bibnamefont
  {Cullity}}\ and\ \bibinfo {author} {\bibfnamefont {C.~D.}\ \bibnamefont
  {Graham}},\ }\href@noop {} {\emph {\bibinfo {title} {Introduction to Magnetic
  Materials}}}\ (\bibinfo  {publisher} {John Wiley \& Sons},\ \bibinfo {year}
  {2005})\BibitemShut {NoStop}%
\bibitem [{\citenamefont {Guinier}\ and\ \citenamefont
  {Fournet}(1955)}]{fournet}%
  \BibitemOpen
  \bibfield  {author} {\bibinfo {author} {\bibfnamefont {A.}~\bibnamefont
  {Guinier}}\ and\ \bibinfo {author} {\bibfnamefont {G.}~\bibnamefont
  {Fournet}},\ }\href@noop {} {\emph {\bibinfo {title} {Small-Angle Scattering
  of X-rays}}}\ (\bibinfo  {publisher} {Wiley},\ \bibinfo {address} {New
  York},\ \bibinfo {year} {1955})\BibitemShut {NoStop}%
\bibitem [{\citenamefont {Porod}(1982)}]{porod}%
  \BibitemOpen
  \bibfield  {author} {\bibinfo {author} {\bibfnamefont {G.}~\bibnamefont
  {Porod}},\ }in\ \href@noop {} {\emph {\bibinfo {booktitle} {Small Angle X-ray
  Scattering}}},\ \bibinfo {editor} {edited by\ \bibinfo {editor}
  {\bibfnamefont {O.}~\bibnamefont {Glatter}}\ and\ \bibinfo {editor}
  {\bibfnamefont {O.}~\bibnamefont {Kratky}}}\ (\bibinfo  {publisher} {Academic
  Press},\ \bibinfo {address} {London},\ \bibinfo {year} {1982})\ pp.\ \bibinfo
  {pages} {17--51}\BibitemShut {NoStop}%
\bibitem [{\citenamefont {Feigin}\ and\ \citenamefont
  {Svergun}(1987)}]{feigin}%
  \BibitemOpen
  \bibfield  {author} {\bibinfo {author} {\bibfnamefont {L.~A.}\ \bibnamefont
  {Feigin}}\ and\ \bibinfo {author} {\bibfnamefont {D.~I.}\ \bibnamefont
  {Svergun}},\ }\href@noop {} {\emph {\bibinfo {title} {Structure Analysis by
  Small-Angle X-Ray and Neutron Scattering}}}\ (\bibinfo  {publisher} {Plenum
  Press},\ \bibinfo {address} {New York},\ \bibinfo {year} {1987})\BibitemShut
  {NoStop}%
\bibitem [{\citenamefont {Michels}\ \emph {et~al.}(2003)\citenamefont
  {Michels}, \citenamefont {Viswanath}, \citenamefont {Barker}, \citenamefont
  {Birringer},\ and\ \citenamefont {Weissm\"uller}}]{michels03prl}%
  \BibitemOpen
  \bibfield  {author} {\bibinfo {author} {\bibfnamefont {A.}~\bibnamefont
  {Michels}}, \bibinfo {author} {\bibfnamefont {R.~N.}\ \bibnamefont
  {Viswanath}}, \bibinfo {author} {\bibfnamefont {J.~G.}\ \bibnamefont
  {Barker}}, \bibinfo {author} {\bibfnamefont {R.}~\bibnamefont {Birringer}}, \
  and\ \bibinfo {author} {\bibfnamefont {J.}~\bibnamefont {Weissm\"uller}},\
  }\href@noop {} {\bibfield  {journal} {\bibinfo  {journal} {Phys. Rev. Lett.}\
  }\textbf {\bibinfo {volume} {91}},\ \bibinfo {pages} {267204} (\bibinfo
  {year} {2003})}\BibitemShut {NoStop}%
\bibitem [{\citenamefont {Weissm\"uller}\ \emph {et~al.}(2004)\citenamefont
  {Weissm\"uller}, \citenamefont {Michels}, \citenamefont {Michels},
  \citenamefont {Wiedenmann}, \citenamefont {Krill~III}, \citenamefont
  {Sauer},\ and\ \citenamefont {Birringer}}]{weissm04a}%
  \BibitemOpen
  \bibfield  {author} {\bibinfo {author} {\bibfnamefont {J.}~\bibnamefont
  {Weissm\"uller}}, \bibinfo {author} {\bibfnamefont {A.}~\bibnamefont
  {Michels}}, \bibinfo {author} {\bibfnamefont {D.}~\bibnamefont {Michels}},
  \bibinfo {author} {\bibfnamefont {A.}~\bibnamefont {Wiedenmann}}, \bibinfo
  {author} {\bibfnamefont {C.~E.}\ \bibnamefont {Krill~III}}, \bibinfo {author}
  {\bibfnamefont {H.~M.}\ \bibnamefont {Sauer}}, \ and\ \bibinfo {author}
  {\bibfnamefont {R.}~\bibnamefont {Birringer}},\ }\href@noop {} {\bibfield
  {journal} {\bibinfo  {journal} {Phys. Rev. B}\ }\textbf {\bibinfo {volume}
  {69}},\ \bibinfo {pages} {054402} (\bibinfo {year} {2004})}\BibitemShut
  {NoStop}%
\bibitem [{\citenamefont {D\"obrich}\ \emph {et~al.}(2012)\citenamefont
  {D\"obrich}, \citenamefont {Kohlbrecher}, \citenamefont {Sharp},
  \citenamefont {Eckerlebe}, \citenamefont {Birringer},\ and\ \citenamefont
  {Michels}}]{dobrichprb2012}%
  \BibitemOpen
  \bibfield  {author} {\bibinfo {author} {\bibfnamefont {F.}~\bibnamefont
  {D\"obrich}}, \bibinfo {author} {\bibfnamefont {J.}~\bibnamefont
  {Kohlbrecher}}, \bibinfo {author} {\bibfnamefont {M.}~\bibnamefont {Sharp}},
  \bibinfo {author} {\bibfnamefont {H.}~\bibnamefont {Eckerlebe}}, \bibinfo
  {author} {\bibfnamefont {R.}~\bibnamefont {Birringer}}, \ and\ \bibinfo
  {author} {\bibfnamefont {A.}~\bibnamefont {Michels}},\ }\href@noop {}
  {\bibfield  {journal} {\bibinfo  {journal} {Phys. Rev. B}\ }\textbf {\bibinfo
  {volume} {85}},\ \bibinfo {pages} {094411} (\bibinfo {year}
  {2012})}\BibitemShut {NoStop}%
\bibitem [{\citenamefont {Romming}\ \emph {et~al.}(2015)\citenamefont
  {Romming}, \citenamefont {Kubetzka}, \citenamefont {Hanneken}, \citenamefont
  {von Bergmann},\ and\ \citenamefont {Wiesendanger}}]{wiesendanger2015}%
  \BibitemOpen
  \bibfield  {author} {\bibinfo {author} {\bibfnamefont {N.}~\bibnamefont
  {Romming}}, \bibinfo {author} {\bibfnamefont {A.}~\bibnamefont {Kubetzka}},
  \bibinfo {author} {\bibfnamefont {C.}~\bibnamefont {Hanneken}}, \bibinfo
  {author} {\bibfnamefont {K.}~\bibnamefont {von Bergmann}}, \ and\ \bibinfo
  {author} {\bibfnamefont {R.}~\bibnamefont {Wiesendanger}},\ }\href@noop {}
  {\bibfield  {journal} {\bibinfo  {journal} {Phys. Rev. Lett.}\ }\textbf
  {\bibinfo {volume} {114}},\ \bibinfo {pages} {177203} (\bibinfo {year}
  {2015})}\BibitemShut {NoStop}%
\bibitem [{\citenamefont {Michels}\ \emph {et~al.}(2000)\citenamefont
  {Michels}, \citenamefont {Weissm\"uller}, \citenamefont {Wiedenmann},
  \citenamefont {Pedersen},\ and\ \citenamefont {Barker}}]{michels00b}%
  \BibitemOpen
  \bibfield  {author} {\bibinfo {author} {\bibfnamefont {A.}~\bibnamefont
  {Michels}}, \bibinfo {author} {\bibfnamefont {J.}~\bibnamefont
  {Weissm\"uller}}, \bibinfo {author} {\bibfnamefont {A.}~\bibnamefont
  {Wiedenmann}}, \bibinfo {author} {\bibfnamefont {J.~S.}\ \bibnamefont
  {Pedersen}}, \ and\ \bibinfo {author} {\bibfnamefont {J.~G.}\ \bibnamefont
  {Barker}},\ }\href@noop {} {\bibfield  {journal} {\bibinfo  {journal}
  {Philos. Mag. Lett.}\ }\textbf {\bibinfo {volume} {80}},\ \bibinfo {pages}
  {785} (\bibinfo {year} {2000})}\BibitemShut {NoStop}%
\end{thebibliography}
\bibliographystyle{apsrev4-1}

\end{document}